\documentclass[aps,floatfix,superscriptaddress,notitlepage,nofootinbib]{revtex4-1}
\usepackage{amsmath,amssymb,amsfonts,graphics,graphicx,dcolumn,bm,enumerate}
\usepackage{comment,natbib,appendix}
\usepackage{multirow,color}
\usepackage{chngpage}
\usepackage{afterpage}
\usepackage{xcolor}
\usepackage{amsthm}
\usepackage{natbib}
\usepackage{hyperref}
\usepackage[margin=1.2in]{geometry}
\usepackage{epstopdf}
\usepackage{float}
\usepackage{amsmath}

\newcommand{\cmp}
{\affiliation{Saha Institute of Nuclear Physics, Kolkata 700064, India.}}

\newcommand{\isi}
{\affiliation{Economic Research Unit, Indian Statistical Institute, Kolkata 700108, India.}}

\newcommand{\raghunathpur}
{\affiliation{Department of Physics, Raghunathpur College, Raghunathpur, Purulia 723133, India.}}

\newcommand{\srm}
{\affiliation{Department of Physics, SRM University - AP,
Department of Computer Science and Engineering, SRM University - AP,
Amaravati, 522240, India.}}

\begin{document}

\title{Signature of maturity in cryptocurrency volatility}

\author{Asim Ghosh}
\email[Email: ]{asimghosh066@gmail.com}
\raghunathpur

\author{Soumyajyoti Biswas}
\email[Email: ]{soumyajyoti.b@srmap.edu.in}
\srm
 
 \author{Bikas K. Chakrabarti }%
 \email[Email: ]{bikask.chakrabarti@saha.ac.in}
 \cmp \isi 

\begin{abstract}
    We study the fluctuations, particularly the inequality of fluctuations, in cryptocurrency prices over the last ten years. We calculate the inequality in the price fluctuations through different measures, such as the Gini and Kolkata indices, and also the $Q$ factor (given by the ratio between the highest value and the average value) of these fluctuations. We compare the results with the equivalent quantities in some of the more prominent national currencies and see that while the fluctuations (or inequalities in such fluctuations) for cryptocurrencies were initially significantly higher than national currencies, over time the fluctuation levels of cryptocurrencies tend towards the levels characteristic of national  currencies. We also compare similar quantities for a few prominent stock prices. 
\end{abstract}

\maketitle

\section{Introduction}
Since its conceptualization in 2008 \cite{nakamoto}, distributed ledger and its application to cryptocurrencies have gathered a lot of attention from financial sectors as well as from data scientists and physicists (see e.g., \cite{btc1,btc2,btc3}). The primary idea that a currency without a central bank backing is revolutionary and consequently gathered a lot a interests and skepticism. From a point of view of purely collective dynamical system showing emergent behavior, therefore, it is an important case study. 

It is known that the fluctuation properties (in terms of prices) of cryptocurrencies have shown a different behavior from regulated currencies \cite{collect}. It is also known that its transaction network, the dynamical properties of such networks and the responses of the prices to world events are far more prominent than regular national currencies \cite{btc4,btc5}. It is important, therefore, to analyse how do these properties change over time and whether it is likely that we would see a behavior, in terms of the price fluctuations, in the cryptocurrencies that are more closer to the other regulated currencies in the future. In this paper, we address these issues through a study in the dynamics of daily price fluctuations in cryptocurrencies and the quantification of the inequality of such fluctuating time series and finally compare those behavior with regulated currencies and some stocks in the markets. The quantification of inequalities are done through first constructing a Lorenz curve $L(p)$, where it denotes, for any fluctuating series of values, the ratio of the sum of $p$ fraction of the smallest values in the series and the sum of all values in the series. First introduced in the context of wealth inequality, it meant that $L(p)$ denoted the ratio of total wealth owned by $p$ fraction of the poorest individuals and the total wealth of the society. Of course, the definition could be extended to the distribution or series of any fluctuating quantity.

\begin{table}[!h]
\centering
\begin{tabular}{|l|r|r|r|r|r|r|}
\hline
\textbf{Sectors} & \textbf{N} & \textbf{Max} & \textbf{Ave} & \textbf{Q} & \textbf{g} & \textbf{k} \\
\hline
agri&70&103373&4264.91&24.24&0.8115&0.8265\\ 
 \hline 
automobile ancillaries&124&383618&26005.47&14.75&0.8006&0.8221\\ 
 \hline 
banks&39&1267323&130499.72&9.71&0.7353&0.7873\\ 
 \hline 
capital goods&122&264822&12256.88&21.61&0.7629&0.7960\\ 
 \hline 
chemicals&157&277290&9819.18&28.24&0.7796&0.8091\\ 
 \hline 
construction materials&69&307821&13044.30&23.60&0.8473&0.8485\\ 
 \hline 
consumer durables&33&69044&10988.30&6.28&0.6781&0.7623\\ 
 \hline 
containers packaging&16&4557&1362.69&3.34&0.5583&0.7074\\ 
 \hline 
diamond jewellery&17&301825&22714.29&13.29&0.8722&0.8890\\ 
 \hline 
diversified&28&523859&37551.25&13.95&0.8182&0.8263\\ 
 \hline 
electricals&32&106567&8511.53&12.52&0.7817&0.8255\\ 
 \hline 
\textcolor{red}{etf} &30&44489&2070.10&21.49&0.9409&0.9380\\ 
 \hline 
finance&148&441607&20962.09&21.07&0.8033&0.8176\\ 
 \hline 
fmcg&61&573606&32885.00&17.44&0.8074&0.8286\\ 
 \hline 
healthcare&140&352042&16673.66&21.11&0.7679&0.8021\\ 
 \hline 
hospitality&33&90744&7681.45&11.81&0.7117&0.7667\\ 
 \hline 
infrastructure&56&497015&21995.62&22.60&0.8726&0.8623\\ 
 \hline 
logistics&49&66468&4362.78&15.24&0.7956&0.8161\\ 
 \hline 
media entertainment&38&30587&2857.61&10.70&0.7832&0.8272\\ 
 \hline 
metals mining&101&295934&22184.81&13.34&0.8336&0.8499\\ 
 \hline 
oil gas&16&1967829&177870.75&11.06&0.8341&0.8473\\ 
 \hline 
paper&20&8634&1297.05&6.66&0.6143&0.7297\\ 
 \hline 
\color{red}{plastic products}&44&74496&4893.23&15.22&0.8403&0.8455\\ 
 \hline 
power&34&348886&44052.32&7.92&0.7400&0.7980\\ 
 \hline 
real estate&68&211911&14558.85&14.56&0.8016&0.8271\\ 
 \hline 
retailing&24&312667&25043.83&12.48&0.8732&0.8793\\ 
 \hline 
 \color{red}{software it services}&141&1378763&28139.64&49.00&0.9141&0.8963\\ 
 \hline 
telecom&24&819744&52046.62&15.75&0.8465&0.8486\\ 
 \hline 
textiles&127&44402&2491.00&17.82&0.7925&0.8275\\ 
 \hline 
\color{red}{trading}&29&363581&15114.00&24.06&0.9276&0.9124\\ 
 \hline 
    \end{tabular}
    \caption{This table shows various sectors of Indian stocks listed on the National Stock Exchange of India. It includes the total number of stocks, the highest market capitalizations ((INR, in Crores), the average market capitalizations ((INR, in Crores), and the $Q$, $g$, and $k$ values for each sector (sectors with unusually high values of the fluctuation inequality indices $g$ and $k$ are indicated in red). The data has been taken from  \url{https://www.moneycontrol.com} during the period June-August, 2024.
    }
 \label{table1}
\end{table}

Having defined the Lorenz function, the common ways of quantifying the inequality would be to calculate: (a) The Gini index \cite{gini}, which refers to the ratio of the area under the Lorenz curve \cite{lor} and that under the equality line (the form of the Lorenz curve, if all values in the series were exactly the same). The Lorenz curve, by definition is bound by $L(0)=0$ and $L(1)=1$ and the equality line is simply a straight line between (0,0) and (1,1). (b) The Kolkata index $k$ \cite{kolkata}, on the other hand, is defined as the the fraction $k$ of the sum of all values in the series that equals $1-k$ fraction of the largest events. In terms of wealth, again, it meant that $1-k$ fraction of the richest individuals own $k$ fraction of the total wealth. Note that it is a generalization of the Pareto's 80-20 law that states 80\% of a society's wealth is owned by 20\% of the richest individuals \cite{pareto}. A lot of analysis have been done using real data on the validity of the Pareto's law and its above mentioned generalization, starting from income inequality assessed through tax returns in the US, shares of individual citations of academic scholars, shares of movie incomes, and so on \cite{banerjee23}. Remarkably, a near
universal pattern of $g = k = 0.84 \pm 0.03$ were seen
in many of these cases. It was conjectured \cite{succ_front} that such a pattern is an outcome of unrestricted competition among `agents' for some `resources'. In a way, the system would self-organize to a point where the inequality among the entities are high, but also bound within a range of about $0.82$ in terms of $g$ and $k$. It was shown later that in generic models of self-organized critical systems, particularly in Bak-Tang-Wiesenfeld and Manna sandpile models that the inequality among the avalanches (responses of the system) would indeed have a near-universal value of about $0.85$ \cite{manna}. Subsequently, such inequalities were calculated in the cases of physical systems (Ising magnets, percolation, fiber bundle model of fracture) and it was shown that systems showing power-law responses would indeed have a value of $g$ and $k$ that are nearly equal close to the critical point, where the fluctuations in the responses are the highest (divergent in the thermodynamic limits, and limited by the system size in finite systems) \cite{soumyaditya}. Also, the value at which the two indices become equal are only weakly dependent on the exponent value of the diverging response function for which the inequality indices are calculated, signalling the origin of the near universal pattern seen in the real data. The crossing of $g$ and $k$, or the approach of either one towards 0.82  could be used as a signal of approaching
critical point or large responses/fluctuations, that was shown
numerically and experimentally \cite{baro24}. Finally, another measure
of inequality used here is (c) the $Q$ factor \cite{q-fact}, which is the ratio between the highest and the average value of signals in a series. In a way it is a background adjusted signal strength of the collective response of a system. It has been shown to be indicative of extraordinary signals from a system near the critical point, in physical as well as social systems. 

In the rest of the paper, we first discuss the result of unrestricted competition across different sectors of the market and show the near-universal inequality pattern among the competing players of these sectors, indicating a self-organized collective behavior. Then, we focus particularly on the cryptocurrency's behavior and show its price fluctuation inequalities and draw comparisons with similar measures for national  currencies.  

\begin{figure}[!tbh]
    \centering
    \includegraphics[width=0.7\textwidth]{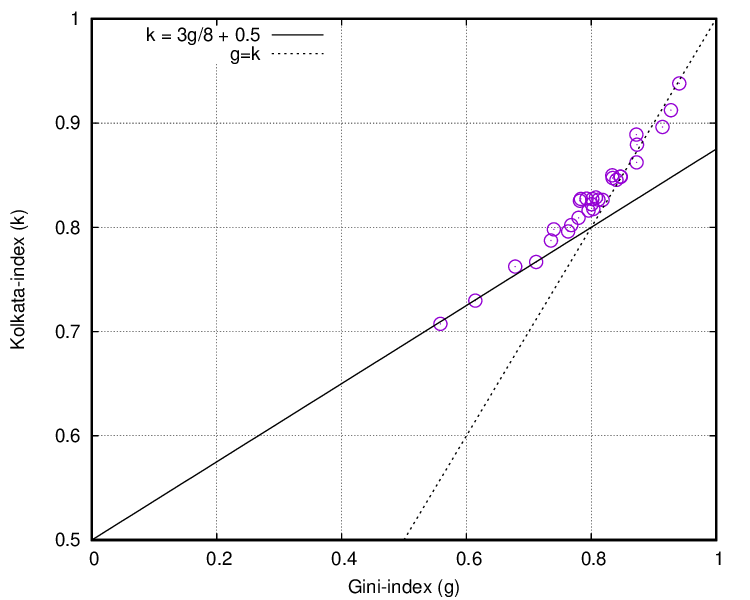}
    \caption{The variation of $k$ with $g$ for different sectors in Indian stocks is shown here. For lower values of $g$ or $k$, the data fits with the line $k = \frac{3}{8}g + \frac{1}{2}$. The data for the plot is taken from Table \ref{table1}.}
    \label{table-g-k.eps}
\end{figure}

\section{Inequality of Market Capitalizations Across Different Sectors}

As discussed above, an indication of an unrestricted competition in markets is a resulting near-universal inequality among the competing agents. In Table \ref{table1}, we show the inequality among the market-caps of different stocks in 27 sectors of the Indian stock market. It is seen that the inequalities of the market-caps, quantified by $g$ and $k$, among the different stocks of various sectors show a near-universal pattern of $g = k \approx 0.82$ (with some exceptions). In rare occasions, the inequalities are significantly higher. The $Q$ value, on the other hand, while showing occurrences of large fluctuations among the market-cap values, do not immediately compare across different sectors.

 In spite of wide variations in the nature of the sectors, the numbers of stocks in all sectors and their market-cap values, the inequalities of the market-caps within individual sectors show near-universal patterns. The near-universal pattern in the market-cap inequalities of stocks in various sectors indicates a self-organized critical state of the market.

Apart from most of the sectors having inequality ($g$ and $k$) close to $0.82$, there is indeed a spread in those values with some sectors having significantly less and some having significantly higher inequality. While an immediate reason is not apparent, when $g$ and $k$ are plotted against each other (see Fig. \ref{table-g-k.eps}) they follow a straight line for smaller values of $g$ until the limit $g=k$, beyond which the two values more or less stay equal.  

A Landau theory like  phenomenological expansion
proposed earlier  that keeps a minimal power
series expansion of the Lorenz function in the form $L(p)=Ap+Bp^2$, which then leading to \cite{joseph22} $k=\frac{1}{2}+\frac{3g}{8}$, which is very close to what is seen in Fig. \ref{table-g-k.eps}. Another non-linear form for the relation between $g$
versus $k$ was also proposed (see \cite{succ_front}). 

However, it is worth emphasising at this point that for the over all state of the market, the inequality of market-cap for the individual stocks within a sector is so adjusted that there is almost a level of universal inequality. The small fraction of sectors that deviate from the near universal region also fall within a very regular relation between $g$ and $k$ that are widely seen elsewhere ( see e.g.,  \cite{banerjee23}). The manifestation of remarkable regularity signal an underlying self-organized critical behavior of the markets that could be immensely helpful in understanding its dynamics. 

\section{Inequality of Daily Price Fluctuation  and Q Value}

Here we focus on the price fluctuations in different cryptocurrencies, some prominent stocks and some national  currencies.

\begin{figure}[!tbh]
    \centering
    \includegraphics[width=1.\linewidth]{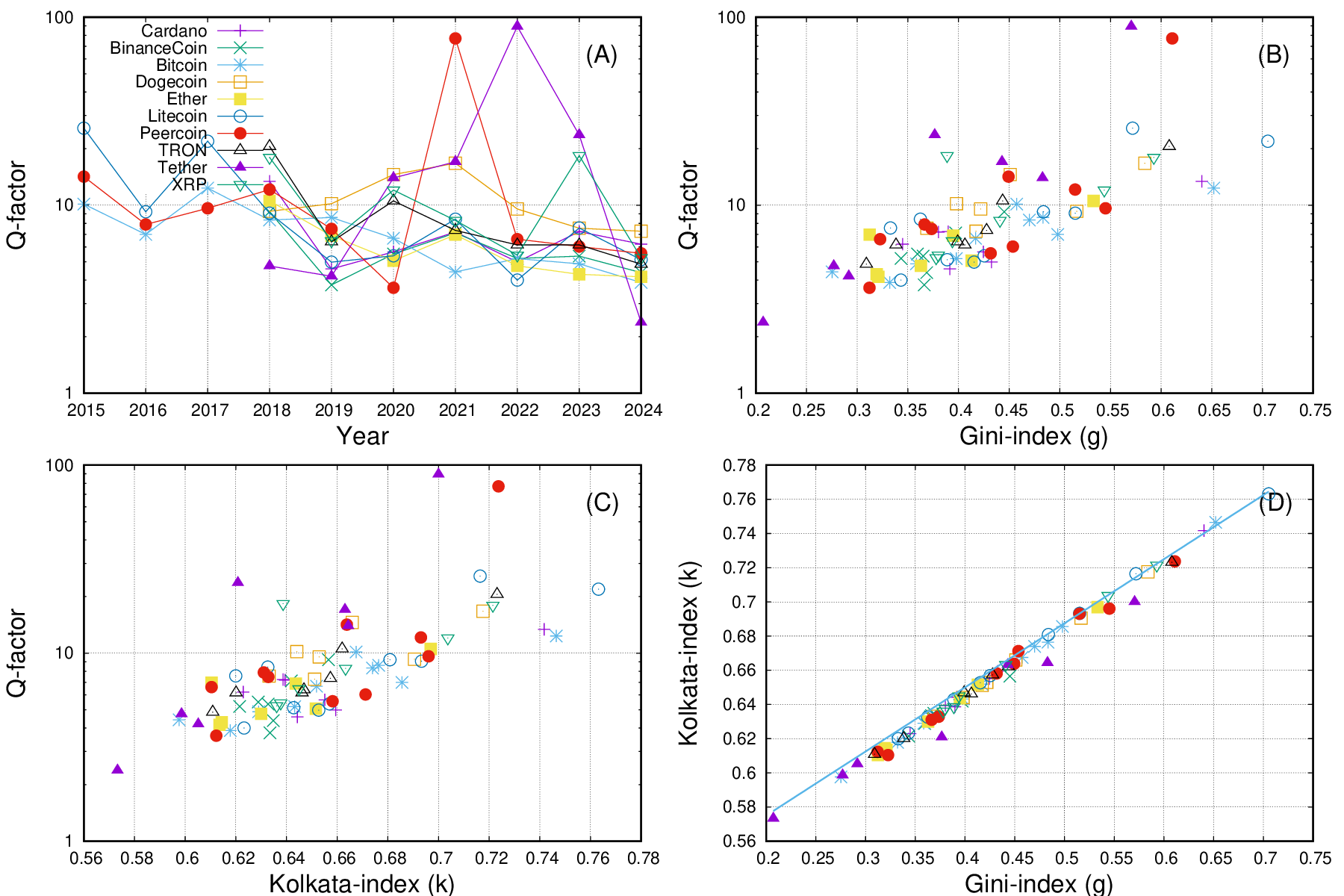}
    \caption{We selected 10 well-known cryptocurrencies and their values in units of US dollars. Daily records of the highest and lowest prices of the cryptocurrencies were collected for each year. The difference between the daily highest and lowest values has been taken, and the $Q$, $g$, and $k$ values have been calculated for each year. (A) The variation of $Q$ over the years is shown. (B) The $Q$ vs. $g$ plot shows a positive correlation. (C) A similar correlation is observed with the $k$-index. (D) The $k$ vs. $g$ plot is compared with the line $k = \frac{3g}{8} + \frac{1}{2}$. The data has been taken from  \url{https://finance.yahoo.com} during the period  June-August, 2024.}
    \label{gkQ_crypto}
\end{figure}
\subsection{Cryptocurrencies}
We take the 10 most prominent cryptocurrencies in terms of their total wealth and look at the daily price fluctuation over a period of about 8 years (from 2015 or later). For each working day in a year, we take the highest ($p_h$) and lowest ($p_l$) prices and find their absolute difference $|p_h-p_l|$. Then for the values obtained for a year, we construct the Lorenz curve $L(p)$ as stated above. Then the inequality indices $g$ and $k$ are calculated respectively, for each year, using $g=1-2\int\limits_0^1L(p)dp$ and solving $1-k=L(k)$. Also, the value of $Q$ for a year is calculated as the ratio of the highest value of price difference in that and the average price difference throughout the year. The results are shown in Fig. \ref{gkQ_crypto}.


The results show that barring a couple of outliers, the $Q$ values have generally decreased over the years for various cryptocurrencies. This suggests a stabilizing behavior in them. We will come back to this point later, while comparing with the behavior seen in national  currencies. 

It is also seen that there is a positive correlation, indeed a possible exponential growth, between the $Q$ values and $g$, $k$ values, as seen in the semi-log plots in Fig. \ref{gkQ_crypto} showing a near straight line trend. This is somewhat expected in the sense that first of all
a high inequality in fluctuation is expected to  be manifested
through different measures.  Also, the $Q$ value in physical
system of percolating clusters \cite{q-fact} have shown almost a
similar behavior as usual response functions, which are
known to grow significantly as $g$ or $k$ approach the
critical point  \cite{succ_front,soumyaditya}. Here, however, the values of $g$ and $k$ do not rise to values high enough where they are expected to be equal. That is why the variation between $g$ and $k$, also shown in Fig. \ref{gkQ_crypto} follow a robust linear trend.


\begin{figure}[!tbh]
    \centering
    \includegraphics[width=0.7\linewidth]{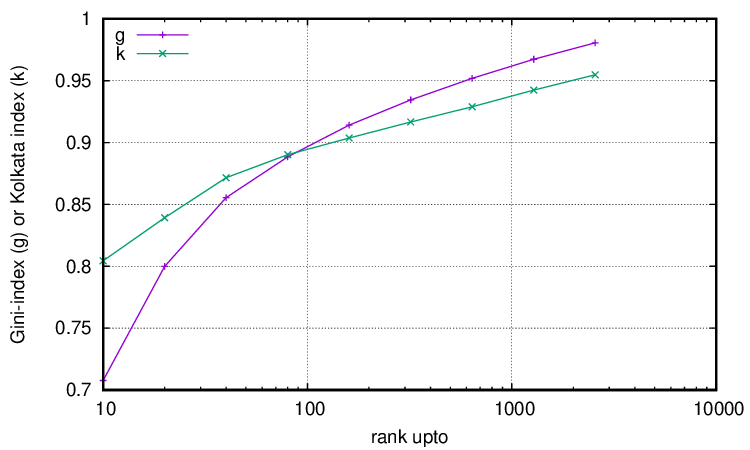}
    \caption{Here, we consider the top 2560 cryptocurrencies ranked by their market capitalization. We determine the $g$ and $k$ values by analyzing the first n-ranked cryptocurrencies. It can be observed that inequalities among the top 120 cryptocurrencies are quite significant. The $g$ value surpasses the $k$ value, with the crossing point approximately at $g=k \approx 0.88$. The data has been taken from  \url{https://www.forbes.com/digital-assets/crypto-prices/?sh=a5edd8c24785} during the period  June-August, 2024.}
    \label{crypto_many}
\end{figure}

Other than the daily price fluctuation characteristics of the individual cryptocurrencies discussed above, we also look at the inequality in the cryptocurrency sector as well, similar to what is reported in Table \ref{table1} for various sectors. The difference for cryptocurrencies is that there are over 2500 cryptocurrencies and many have very little market-cap. Therefore, the overall inequality is extremely high. In Fig. \ref{crypto_many}, we show the inequality ($g$, $k$) of the cryptocurrency wealth values for the top $n$-ranked cryptocurrencies, as a function of $n$. The values of $g$ and $k$ cross as $n$ is increased, and the crossing is seen around $0.82$. 

\subsection{International stocks}
Given that cryptocurrencies are not regulated by a central bank, it is worthwhile to compare its behavior with that of stock prices. Of course, there is no regulation in stock prices of private companies. Therefore, we measure the inequality of daily price fluctuations  in 10 prominent international using the same methods mentioned above, for a period of about 50 years. 

\begin{figure}[!tbh]
    \centering
    \includegraphics[width=1.\textwidth]{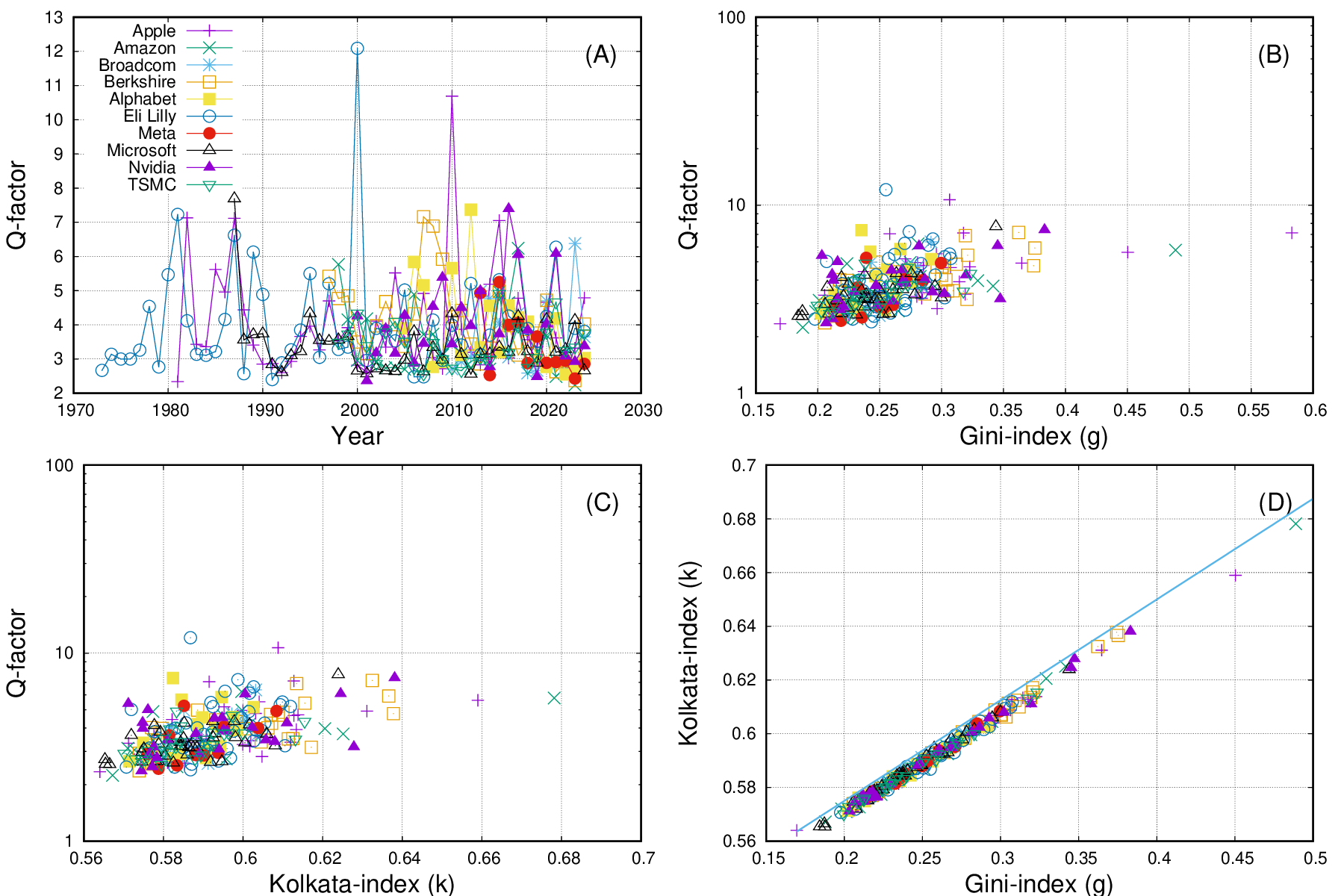}
    \caption{We selected 10 well-known stocks and their values in units of US dollars. Daily records of the highest and lowest prices of the stocks were collected for each year. The difference between the daily highest and lowest values has been taken, and the $Q$, $g$, and $k$ values have been calculated for each year. (A) The variation of $Q$ over the years is shown. (B) The $Q$ vs. $g$ plot shows a positive correlation. (C) A similar correlation is observed with the $k$-index. (D) The $k$ vs. $g$ plot is compared with the line $k = \frac{3g}{8} + \frac{1}{2}$. The data has been taken from  \url{https://finance.yahoo.com} during the period  June-August, 2024.}
    \label{gkQ_int_stocks}
\end{figure}

The results of the international stock price inequalities are shown in Fig. \ref{gkQ_int_stocks}. The $Q$ values show a much higher range of values than what are seen for cryptocurrencies. This is the first indication that in spite of the intuition that unregulated cryptocurrencies might behave similar to the stock prices, in effect they are somewhat different. The positive correlations between $Q$ and $g$, $k$ still exist, albeit with less prominence. The linearity between $g$ vs $k$ plot, however, robustly appears here as well. 

\subsection{Indian stocks}
\begin{figure}[!tbh]
    \centering
    \includegraphics[width=\textwidth]{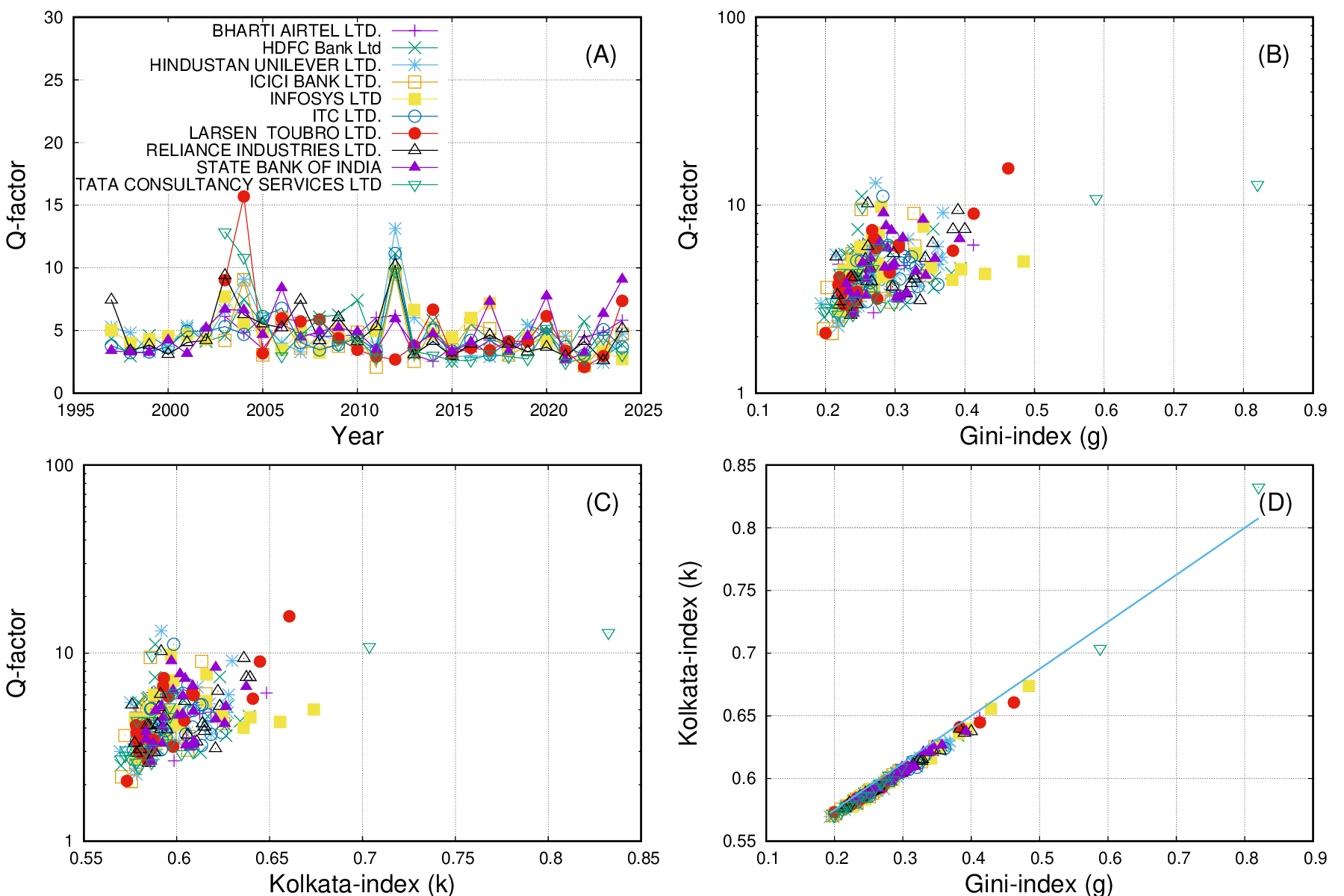}
    \caption{We selected 10 well-known stocks in the Indian market. Daily records of the highest and lowest prices of the stocks (in units of Indian rupee) were collected for each year. The difference between the daily highest and lowest values has been taken, and the $Q$, $g$, and $k$ values have been calculated for each year. (A) The variation of $Q$ over the years is shown. (B) The $Q$ vs. $g$ plot shows a positive correlation. (C) A similar correlation is observed with the $k$-index. (D) The $k$ vs. $g$ plot is compared with the line $k = \frac{3g}{8} + \frac{1}{2}$. The data has been taken from  \url{https://finance.yahoo.com} during the period  June-August, 2024.}
    \label{gkQ_ind_stocks}
\end{figure}

Here we do the same analysis of price fluctuation inequalities, but with 10 prominent stocks from India. Perhaps not surprisingly, we see a similar range of values for $Q$ and similar relations for $g$ and $k$ (see Fig. \ref{gkQ_ind_stocks}). 
An interesting point to note here that given that the stocks from the same country, national events are expected to be mirrored in the stock prices and their fluctuations. A tendency to peak in Indian national election years seems to be  present.

\subsection{National  currencies}
As mentioned before, we need to compare the fluctuation characteristics of cryptocurrencies with national  currencies that are backed up by central banks. We take some of the prominent national currencies and measure their daily price fluctuations as before. Here we take the US dollar as the reference currency i.e., all prices are measured in terms of USD, hence it is absent from the plots.  
\begin{figure}[!tbh]
    \centering
    \includegraphics[width=1.\textwidth]{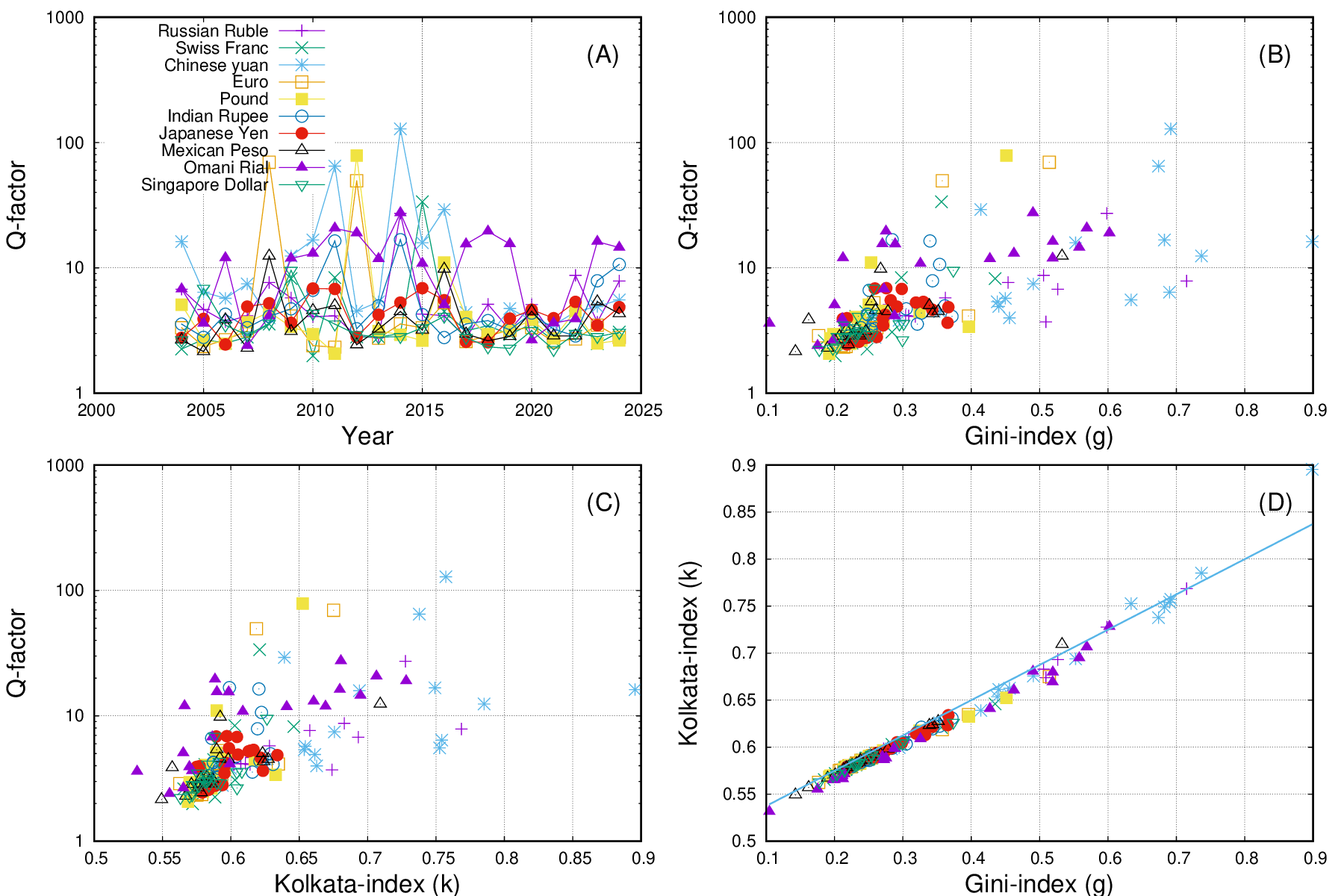}
    \caption{We selected 10 well-known currencies and compared their values against the US dollar (taken as the reference currency). Daily records of the highest and lowest prices were collected from 2004 to 2024. The quantity used for analysis is the difference between the daily highest and lowest values. The $Q$, $g$, and $k$ values are calculated for each year. (A) The variation of $Q$ over the years is shown. (B) The $Q$ vs. $g$ plot shows a positive correlation. (C) A similar correlation is observed with the $k$-index. (D) The $k$ vs. $g$ plot is compared with the line $k = \frac{3g}{8} + \frac{1}{2}$. The data has been taken from  \url{https://finance.yahoo.com} during the period June-August, 2024.
}
    \label{gkQ_currency}
\end{figure}

Fig. \ref{gkQ_currency} shows the inequality of the daily price fluctuations for different currencies. It is interesting to note that the $Q$ values here are closer to the ones seen for the cryptocurrencies than in the cases of various stock prices. Indeed, by comparing Fig. \ref{gkQ_crypto} and Fig. \ref{gkQ_currency}, we see that the $Q$ values for the cryptocurrencies have started off from a higher range in their initial stages, but over time show a tendency towards approaching the values seen in national  currencies rather than the stock prices studied here. This is the most interesting property of the cryptocurrency's price fluctuations characteristics seen here. Even though these are not regulated by any central bank, their price fluctuation characteristics tend towards those seen for other regulated currencies rather than unregulated stock prices. 
 
\section{Discussions and conclusions}
The overall tendency of a multi-component interacting system to manifest emergent collective behavior is well studied in self-organized critical phenomena. Without a fine tuning of a driving parameter to a finite value, the system sets itself in a way that results in a diverging correlation and often an efficient point of operation e.g. in human brain \cite{brain}. The diverging correlation can also be manifested in terms of the inequalities of the response functions/ fluctuations in such systems. The inequality indices, used to quantify the inequality of the fluctuations of responses, are then known to show near-universal signals in various social and physical systems (see e.g.,  \cite{banerjee23,succ_front,manna}). 

Here we looked at the fluctuation properties of cryptocurrencies and tried to identify its dynamical tendencies with other national currencies as well as some prominent stocks in the market. Particularly, we have seen that market-caps within various sectors show a near-universal inequality (see Table \ref{table1}) that are reminiscent of what was observed \cite{manna} for the response functions (avalanches) in self-organized critical systems. This suggests a self-organized critical state of the market as a whole. 

Then we looked at the inequalities of daily price fluctuations in various cryptocurrencies, prominent stocks and prominent national currencies, all estimated against US dollar (of the corresponding year) as a reference. Some observations of regularity are true  across these cases: In all cases $g$ and $k$ show a linear relationship ($k = \frac{1}{2} + \frac{3}{8}g$, as obtained earlier from a Landau-like expansion of the Lorenz function \cite{joseph22}), particularly towards the smaller values of $g$ (see Fig. 2D for cryptocurrencies, Fig. 4D for international stocks, Fig. 5D for Indian stocks, and Fig. 6D for some national currencies). The inequality indices $g$ and  $k$ show a positive correlation with the other measure of inequality, the $Q$ factor (see Figs. 2B and 2C for cryptocurrencies, Figs. 4B and  4C for international stocks,
 Figs. 5B and 5C for Indian stocks,  and Figs. 6B and 6C for some  national currencies.  What is remarkable, however, is the dynamical behavior of $Q$ for cryptocurrencies. Unlike the intuitive expectation of varying like unregulated stocks, the cryptocurrencies, at least in terms of their $Q$ values of daily price fluctuations, behave more like national currencies, as time progresses (though cryptocurrencies do not have any central bank for over-all control, as in the cases of national currencies). This signals a maturity in some of the prominent cryptocurrencies and the trend towards stable currencies.

\section*{acknowledgement}
This paper is dedicated to the memory of our
colleague Prof. Manipushpak Mitra. BKC is grateful to the Indian
National Science Academy for a Senior Scientist Research Grant.

\end{document}